\documentclass[a4paper, 9pt, twocolumn]{article}
\usepackage{geometry}                
\usepackage{graphicx}
\usepackage{amssymb}
\usepackage{amsthm}
\usepackage{epstopdf}
\usepackage{mathtools}
\usepackage[shortlabels]{enumitem}
\usepackage{dsfont}
\usepackage{cuted}
\usepackage{sidecap}
\usepackage[usenames,dvipsnames]{color}
\usepackage[thinspace,mediumqspace,Grey,squaren]{SIunits}
\usepackage{bbm}
\usepackage[T1]{fontenc}
\usepackage[utf8]{inputenc}
\usepackage{authblk}
\usepackage[font=small,labelfont=bf ,textfont={small, sf}]{caption}
\usepackage{titlesec}
\usepackage{url}
\usepackage{hyperref}
\usepackage{algorithm}
\usepackage{algorithmic}
\titleformat{\section}{\normalfont \Large \bfseries}{\thesection }{7pt}{}{}
\titleformat{\subsection}{\normalfont \large \bfseries}{\thesubsection }{7pt}{}{}

\begin{document}
\title{Moment-based analysis of biochemical networks in a heterogeneous population of communicating cells} 

\author[1,2]{David T. Gonzales}
\author[1]{T-Y Dora Tang}
\author[1,2,3]{Christoph Zechner}
\affil[1]{Max Planck Institute of Molecular Cell Biology and Genetics, 01307 Dresden, Germany}
\affil[2]{Center for Systems Biology Dresden, 01307 Dresden, Germany}
\affil[3]{Correspondence to: zechner@mpi-cbg.de}
\date{}
\maketitle
\newpage{}

{\noindent \sf \fontsize{10}{0} \selectfont \textbf{
Cells can utilize chemical communication to exchange information and coordinate their behavior in the presence of noise. Communication can reduce noise to shape a collective response, or amplify noise to generate distinct phenotypic subpopulations. Here we discuss a moment-based approach to study how cell-cell communication affects noise in biochemical networks that arises from both intrinsic and extrinsic sources. We derive a system of approximate differential equations that captures lower-order moments of a population of cells, which communicate by secreting and sensing a diffusing molecule. Since the number of obtained equations grows combinatorially with number of considered cells, we employ a previously proposed model reduction technique, which exploits symmetries in the underlying moment dynamics. Importantly, the number of equations obtained in this way is independent of the number of considered cells such that the method scales to arbitrary population sizes. Based on this approach, we study how cell-cell communication affects population variability in several biochemical networks. Moreover, we analyze the accuracy and computational efficiency of the moment-based approximation by comparing it with moments obtained from stochastic simulations. }}

\section{Introduction}
In recent years, significant progress has been made in developing computational models and algorithms to study stochastic biochemical networks inside living cells. Most commonly, these models are based on the chemical master equation (CME), whose solution provides a time-dependent probability distribution over molecular concentrations \cite{gillespie1992}. CME-based models can account for the discrete and random nature of biochemical reactions (intrinsic noise) as well as population heterogeneity stemming from differences in each cell's microenvironment (extrinsic variability) \cite{elowitz2002}. The computational analysis of the CME is challenging but several efficient numerical techniques have been proposed in the past, including stochastic simulation algorithms \cite{gillespie1977}, moment-based methods \cite{singh2006, zechner2012} and combinations thereof \cite{salis2005, duso2018, ganguly2015}.

However, the majority of existing approaches to study noise in cell populations rely on the assumption that individual cells act independently of each other. More concretely, each cell's dynamics is considered to be an independent and identically distributed realization of the same stochastic process. Evidently, this assumption is violated in systems where cells communicate with one another to coordinate their behavior \cite{smith2018,youk2014}. Typical examples include quorum-sensing systems in bacterial colonies \cite{stevens1994}, or paracrine communication in higher organisms \cite{handly2015}. Understanding the interplay between cell-cell communication and the stochastic behavior of individual cells is an important challenge and demands for suitable mathematical approaches. However, extending standard techniques to account for cell-cell communication leads to computational difficulties, because the dimensionality of the resulting models increases with the number of cells in a population.

Recently, first attempts have been made to develop more tractable stochastic models of systems of communicating cells. In \cite{smith2018}, for instance, the authors use a moment-based method to study how neighbour-neighbour-coupling affects concentration fluctuations in a tissue. A related approach has been proposed in \cite{batmanov2012} to study community effects in cells that interact with each other by secreting and sensing certain signalling molecules. In particular, the authors show how the dimensionality of the considered model can be dramatically reduced by exploiting certain symmetries in the governing equations. However, both approaches account exclusively for intrinsic noise, whereas extrinsic sources have not been considered. In \cite{boada2017}, the authors study how chemical communication via a quorum sensing molecule affects intrinsic and extrinsic noise in cell communities. To obtain tractable simulations, they used a tailored approach that combines stochastic simulations with a quasi-steady state approximation to eliminate fast variables from the model.

In this article, we discuss a moment-based approach to study noise in communicating cells that arises from both intrinsic and extrinsic sources. In particular, we focus on a secrete-and-sense model \cite{youk2014, batmanov2012}, where individual cells can secrete a signalling molecule to the extracellular environment, which in turn can be sensed by other cells in the population. We first derive a system of differential equations that captures lower order moments of the population. Since the number of obtained equations grows combinatorially with number of considered cells, we perform a symmetry-based model reduction based on the work of Batmanov et al. (2012) \cite{batmanov2012}. The dimensionality of the reduced model is independent of the population size, which makes it computationally very efficient. We employ this approach to study how noise is affected by cell-to-cell communication in a few biochemical systems.

The rest of the paper is structured as follows. A general model of biochemical networks in chemically interacting cell populations is presented in Section \ref{model}. In Section \ref{moments}, we provide a stochastic description of this model based on the CME and derive a general equation for its moment dynamics. In Section \ref{reduction}, we show how the number of moment equations can be reduced by exploiting symmetries of the model. Finally, we apply our approach to analyze several biochemical networks in Section \ref{cases}.

\section{Reaction networks in a population of chemically interacting cells}\label{model}
\begin{figure}[h]
\centering
\includegraphics[width=0.98\columnwidth]{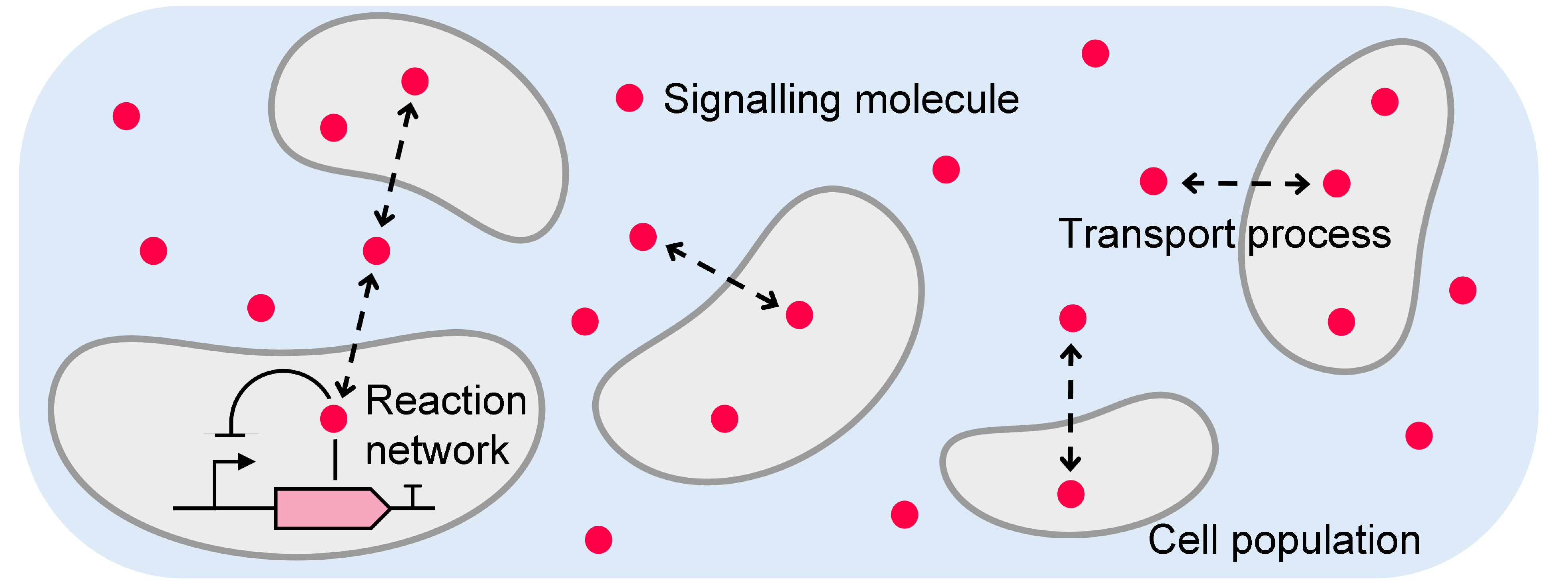}
\vspace*{-1mm}
\caption{Secrete-and-sense model of a cell population. Each cell's dynamics is described by the same reaction network, whereas one of the reactants serves as a signalling molecule. The latter can diffuse to the homogeneous extracellular environment from where it can be sense by any of the cells.}
\label{fig:cells}
\end{figure}

We consider a population of $N$ genetically identical cells that communicate with each other through a diffusing signalling molecule (Fig. \ref{fig:cells}). Each individual cell $i$ is associated with an identical set of $S$ chemical species $\mathrm{X}_{i, 1}, \ldots, \mathrm{X}_{i, S}$ that interact with one another via $M$ biochemical reaction channels. Without loss of generality, we consider the first $S-1$ species to be confined to the intracellular environment of cell $i$. The $S$th species corresponds to the signalling molecule, which can shuttle between the intra- and extracellular environment through transport reactions. For simplicity, we consider the external environment to be well-mixed, such that the import of signalling molecules into any cell $i$ does not depend on the spatial configuration of the system. In total, the system can be described by a reaction network
\begin{equation} \label{rxn_1}
    \begin{split}
    \sum_{k=1}^{S}\alpha_{j,k}\mathrm{X}_{i,k}&\xrightarrow{}\sum_{k=1}^{S}\beta_{j,k}\mathrm{X}_{i,k} \\
    \mathrm{X}_{i,S} &\xleftrightharpoons[]{} \mathrm{X}_{E},
    \end{split}
\end{equation}
for $i=1, \ldots, N$ and $j=1, \ldots, M-2$. In (\ref{rxn_1}), $\alpha_{j,k}$ and $\beta_{j,k}$ correspond to the reactant and product coefficients of the respective reaction. The species $\mathrm{X}_{E}$ denotes the signalling molecule in the external environment. 

In total, the network comprises $N S + 1$ chemical species and $N M$ chemical reactions. We define by $X_i(t) = (X_{i, 1}(t), \ldots, X_{i, S}(t))$ the state of cell $i$, which collects the copy numbers of all species associated with this cell at time $t$. The state of the overall system is then given by $X(t) = (X_1(t), \ldots, X_N(t), X_E(t))$, with $X_E(t)$ as the number of signalling molecules in the external environment. Moreover, we denote by $\nu_{i, k} \in \mathbb{Z}^{N S + 1}$ the stoichiometric change vector associated with the $k$th reaction channel of cell $i$ consistent with reaction network (\ref{rxn_1}).

\section{Moment dynamics of heterogeneous cell communities} \label{moments}

If both the intra- and extracellular environments are well-mixed, we can describe the state $X(t)$ as a continuous-time Markov chain, whose state probability distribution $P(x, t) = P(X(t)=x)$ admits a master equation of the form
\begin{equation} \label{CME}
\begin{split}
\frac{\mathrm{d}P(x,t)}{\mathrm{d}t}&=\sum_{i=1}^{N}\sum_{j=1}^{M}[a_{i,j}(x-\nu_{i,j}, c_{i, j})P(x-\nu_{i,j},t)\\
&\qquad\qquad\qquad\qquad\quad -a_{i,j}(x, c_{i, j})P(x,t)].
\end{split}
\end{equation}
In (\ref{CME}), the function $a_{i,j}$ is the reaction propensity associated with reaction $j$ in cell $i$. Throughout this article, we consider the propensities to obey the law of mass action such that $a_{i,j}(x, c_{i, j}) = c_{i, j} g_{j}(x_i, x_E)$ with $x_i$ as the part of the state vector associated with cell $i$, $x_E$ as the abundance of the signalling molecule in the environment, $c_{i, j} \in \mathbb{R}^+$ as a stochastic rate constant, and $g_{j}$ as a polynomial. Note that while $g_{j}$ is identical for all cells, we allow the individual rate constants $c_{i, j}$ to vary across the population. This provides a means to account for extrinsic sources of cell-to-cell variability \cite{zechner2012}. We consider the reaction rate constants for each cell $i$ to be independent random vectors $C_i = (C_{i, 1}, \ldots, C_{i, M})$ drawn from a common probability distribution $C_i \sim p_c(\cdot)$ for $i=1,\ldots, N$. Note that  deterministic (non-varying) reaction rates can be accounted for by letting $p_c$ be a Dirac measure with respect to this parameter. With $C=(C_1,\ldots, C_N)$, we can formulate a master equation for the conditional distribution $P(x, t \mid c) = P(X(t) = x \mid C = c)$, i.e.,
\begin{equation}
\begin{split}
\frac{\mathrm{d}P(x,t \mid c)}{\mathrm{d}t}&=\sum_{i=1}^{N}\sum_{j=1}^{M}[a_{i,j}(x-\nu_{i,j}, c_{i, j})P(x-\nu_{i,j},t \mid c)\\
&\qquad\qquad\qquad-a_{i,j}(x, c_{i, j})P(x,t \mid c)].
\label{condCME}
\end{split}
\end{equation}

\subsection{Moment dynamics}
To analyze the cell community model, we resort to a moment-based approach, which provides a lower-dimensional description of the population and its heterogeneity. More precisely, we seek for the population moments

\begin{equation}
\begin{split}
    \langle \phi(X,C) \rangle &= \big\langle \langle \phi(X,C) \mid C \rangle \big\rangle\\ &=\Bigg\langle \sum_{x \in \mathcal{X}} \phi(x,C) P(x, t \mid C) \Bigg\rangle,
    \label{phiMoment}
\end{split}
\end{equation}
with $\phi: (x,c) \rightarrow \mathbb{R}$ as a monomial in $(x,c)$ and $\mathcal{X}$ as the domain of $X(t)$. Note that we omit the dependency of the moments on time for the sake of a compact notation. In order to derive a differential equation for the time evolution of $\langle \phi(X,C) \rangle$, we calculate the derivative of (\ref{phiMoment}) and insert the r.h.s. of (\ref{condCME})
\begin{equation}
\small
\begin{split}
    &\frac{\mathrm{d}\langle \phi(X,C) \rangle}{\mathrm{d}t} = \Bigg\langle\sum_{x\in\mathcal{X}} \phi(x,C) \sum_{i=1}^{N}\sum_{j=1}^{M}[a_{i,j}(x-\nu_{i,j}, C_{i, j}) \\
    & \qquad P(x-\nu_{i,j},t \mid C)-a_{i,j}(x, C_{i, j})P(x,t \mid C)] \Bigg\rangle.
    \label{phiMomentEQ}
    \end{split}
\end{equation}
Using a change of variable, it is straightforward to show that (\ref{phiMomentEQ}) simplifies to
\begin{equation}
\small
\begin{split}
    &\frac{\mathrm{d}\langle \phi(X,C) \rangle}{\mathrm{d}t} = \Bigg\langle\sum_{i=1}^{N}\sum_{j=1}^{M}\Big \langle \phi(X+\nu_{i, j},C) a_{i,j}(X, C_{i, j}) \mid C \Big \rangle\\
&\qquad\qquad\qquad - \Big  \langle \phi(X,C) a_{i,j}(X, C_{i, j}) \mid C \Big \rangle \Bigg\rangle,
    \label{phiMomentEQ2}
    \end{split}
\end{equation}
where the inner brackets denote expectations conditionally on the random parameters $C$. Now, using double expectations, we obtain
\begin{equation}
\small
\begin{split}
    \frac{\mathrm{d}\langle \phi(X,C) \rangle}{\mathrm{d}t}  &= \sum_{i=1}^{N}\sum_{j=1}^{M}\Big \langle \phi(X+\nu_{i, j},C) a_{i,j}(X, C_{i, j}) \Big \rangle\\
&\qquad\qquad\quad - \Big  \langle \phi(X,C) a_{i,j}(X, C_{i, j})\Big \rangle.
    \label{phiMomentEQ3}
    \end{split}
\end{equation}
Eq. (\ref{phiMomentEQ3}) describes the time evolution of moments and cross moments for a heterogeneous population of secrete-and-sense cells. It can thus be seen as an extension of the moment equations derived in \cite{zechner2012} to account for cell-to-cell communication. Indeed, if we set the transport rates to zero, all cells in the population become independent of each other such that the two approaches yield equivalent solutions.

Note that depending on the details of the system, the moment equations from (\ref{phiMomentEQ3}) may not be closed. This is the case in the presence of second-order reactions, or even first-order reactions if their rate constants are randomly distributed across the population. Moment-closure approximation (MA) techniques provide a popular means to address this problem by imposing certain assumptions on the underlying state probability distribution \cite{singh2006, zechner2012, schnoerr2015}. More precisely, this allows us to replace the higher order moments that appear on the r.h.s. of (\ref{phiMomentEQ3}) by functions of the lower order moments. These functions are referred to as \textit{closure functions} and their particular form depends on the distributional assumption we make. Popular choices include the normal \cite{schnoerr2015} and lognormal \cite{singh2006} closure functions and we will adopt those in the present study. To check the accuracy of our MAs in this study, we compare them with moments calculated using the Stochastic Simulation Algorithm (SSA) \cite{gillespie1977}. Throughout this article, we consider moments of up to second order and replace all third-order moments that appear on the r.h.s. of (\ref{phiMomentEQ3}) using MA functions provided in Table \ref{closure}.

\begin{table}[h!]
    \centering
    \footnotesize
    \caption{Normal \& lognormal closure functions of order three.}
    \begin{tabular}{cc}
    \hline
    \noalign{\vskip 1mm} 
    MA & $\langle X_1 X_2 X_3 \rangle$ \\
    \noalign{\vskip 1mm}
    \hline
    \noalign{\vskip 1mm}  
    Normal &  $\langle X_1 \rangle \langle X_2 X_3 \rangle + \langle X_2 \rangle \langle X_1 X_3 \rangle$\\
     & $ + \langle X_3 \rangle \langle X_1 X_2 \rangle - 2\langle X_1 \rangle \langle X_2 \rangle \langle X_3 \rangle$\\
    \noalign{\vskip 1mm}
    Lognormal & $\dfrac{\langle X_1 X_2 \rangle \langle X_2 X_3 \rangle \langle X_1 X_3 \rangle}{\langle X_1 \rangle \langle X_2 \rangle \langle X_3 \rangle}$\\
    \noalign{\vskip 1mm}
    \hline
    \end{tabular}
    \label{closure}
\end{table}

\section{Symmetry-based model reduction} While eq. (\ref{phiMomentEQ3}) provides a more tractable description of the cell community than (\ref{CME}), its dimensionality still scales combinatorially with the number of considered cells $N$. In case we consider all first and second order moments, the number of equations is given by
\begin{equation} \label{het_scale}
K_{eq}=2(NS+1) + \binom{NS+NM'+1}{2}-\binom{NM'}{2},
\end{equation}
where $M'$ is the number of rate constants among the $M$ reactions that vary from cell to cell. The first term in (\ref{het_scale}) accounts for 1st and 2nd order moments of the chemical species, and the remaining terms account for all cross-moments of species and rate constants that change over time. 

The number of first and second order moment equations scales quadratically with $N$, which limits the above approach to relatively small population sizes. However, the moment system can be reduced substantially by taking into account the symmetries of the considered model as has been proposed in \cite{batmanov2012}. More precisely, if we consider all initial cell states $X_i(0)$ to be identically distributed, the moment dynamics of each cell will be equivalent and indistinguishable for all times $t>0$ such that $\langle X_{i,k} \rangle = \langle X_{j,k}\rangle$,  $\langle X_{i,k}^2 \rangle = \langle X_{j,k}^2\rangle$, $\langle X_{i,k}X_{j,l} \rangle = \langle X_{m,k}X_{n,l} \rangle$,  and $\langle X_{i,k}X_E \rangle = \langle X_{j,k}X_E \rangle$ for any $i \neq j$, $m \neq n$, $k$ and $l$. Consequently, we can obtain a reduced model by considering only the moments up to the second order and cross-moments of the states of any two reference cells $X_i(t)$ and $X_j(t)$ as well as the amount of signalling molecules in the external environment $X_E(t)$. The resulting set of equations can be reduced further by eliminating all moments associated with one of the two cells $j$ (e.g., $\langle X_{j,k} \rangle$ or $\langle X_{j,k}X_E \rangle$) since those are identical to the corresponding moments of cell $i$.
In total, the required number of equations is then given by
\begin{equation} \label{reduced_scale}
\begin{split}
\hat{K}_{eq}=2(S+1)+&\binom{S+1+M'}{2}\\
\qquad \qquad \qquad \qquad &+\binom{S+M'}{2}-2\binom{M'}{2}+S,
\end{split}
\end{equation}
which is thus independent of the population size.

\subsection{Illustrative example}
Consider the toy model
\begin{equation}
\begin{gathered}
\begin{aligned}[c]
\mathrm{X}_{i,1} \xrightarrow{C_{i,1}} \mathrm{X}_{i,2} \\
\end{aligned}
\qquad\quad
\begin{aligned}[c]
\mathrm{X}_{i,2} \xleftrightharpoons[c_{3}]{c_{2}} \mathrm{X}_E,
\end{aligned}
\end{gathered}
\end{equation}

with $i=1, \ldots, N$. We consider the rate $C_{i, 1}$ to be randomly distributed across the population, while $c_2$ and $c_3$ are identical in all cells.

To demonstrate how the original system can be reduced based on symmetries, we distinguish between two different cases. The first case concerns equations that are the same in the original and reduced model. These are equations of the moments that do not involve the signalling molecule in the external environment such as $\langle X_{i,1} \rangle$, $\langle X_{i,2} \rangle$, or $\langle X_{i,1}X_{i,2} \rangle$. For example, in both the original and reduced models, the expectation of species $X_{i,2}$ satisfies
\begin{equation} \label{unreduced_x12}
\frac{\mathrm{d}\langle X_{i,2}\rangle}{\mathrm{d}t}= \langle C_{i,1}X_{i,1}\rangle + c_{2} \langle X_E \rangle - c_{3} \langle X_{i,2} \rangle.
\end{equation}
The dynamics of moments and cross-moments involving the external signalling molecule depend on \textit{all} cells in the population due to the transport reactions. In the case of $\langle X_{i,2}X_E \rangle$, the original equation is:
\begin{equation} \label{unreduced_x12x4}
\begin{split}
\frac{\mathrm{d}\langle X_{i,2}X_E\rangle}{\mathrm{d}t} &= \langle C_{i,1}X_{i,1}X_E \rangle + c_{2}\langle X_E^2\rangle - c_{2}\langle X_E\rangle \\
&\hspace{-1em} + c_{3}\langle X_{i,2}^2 \rangle - c_{3}\langle X_{i,2}X_E \rangle - c_{3}\langle X_{i,2}\rangle \\
&\hspace{-1em} + \sum_{k\neq i} c_{3}\langle X_{k,2}X_{i,2}\rangle - c_{2} N \langle X_{i,2} X_E  \rangle.
\end{split}
\end{equation}
The terms in the sum capture the dependencies between two cells $k$ and $i$. 
Since all terms in the sum are identical due to the symmetry of the population, they can be replaced by the contribution of any cell $j\neq i$ multiplied by $(N-1)$,
\begin{equation} \label{reduced_x12x4}
\begin{split}
\frac{\mathrm{d}\langle X_{i,2}X_E\rangle}{\mathrm{d}t} &= \langle C_{i,1}X_{i,1}X_E \rangle + c_{2}\langle X_E^2\rangle- c_{2}\langle X_E\rangle \\
&\hspace{-1em} + c_{3}\langle X_{i,2}^2 \rangle - c_{3}\langle X_{i,2}X_E \rangle - c_{3}\langle X_{i,2}\rangle \\
&\hspace{-1em} + c_{3} (N-1)\langle X_{j,2}X_{i,2}\rangle - c_{2} N \langle X_{i,2} X_E \rangle.
\end{split}
\end{equation}
Therefore, if we perform analogous manipulations for all other moments and cross-moments involving the signalling molecule (e.g., $\langle X_E \rangle$, $\langle X_{i,1} X_E \rangle$) and eliminate all remaining redundancies between cell $i$ and $j$, we arrive at a system of $17$ coupled differential equations, independent of the population size $N$. For further information on the symmetry-based model reduction approach, the reader may refer to \cite{batmanov2012}.

\section{Case studies} \label{cases}
In this section, we use the described moment-based approach to study how cell-cell communication affects noise in different biochemical networks. Python scripts used for this study are available in \url{github.com/zechnerlab/CommunityMoments}. Moment equations were solved using a SciPy numerical solver (solve\_ivp). SSA sample paths from Gillespie's Direct Method were simulated using Tellurium \cite{choi2018} and used to test the accuracy of the moment-based approach. For the sake of a compact notation, molecular species are assigned different letters and reaction rate parameters are assigned letters with superscripts.

\subsection{Birth-death process}
As a first example, we study a birth-death process in a heterogeneous population of interacting cells, i.e.,
\begin{equation} \label{birthdeath}
\begin{gathered}
    \begin{aligned}
    \emptyset \xrightarrow{C^b_i} \mathrm{P}_i \xrightarrow{C^d_i} \emptyset \\
    \end{aligned}
    \qquad\qquad
    \begin{aligned}[c]
    \mathrm{P}_i \xleftrightharpoons[c^t]{c^t} \mathrm{Q},
    \end{aligned}
\end{gathered}
\end{equation}
for $i=1, \ldots, N$. Here, the birth and death reaction rate constants are independent random variables as indicated by capital letters $C^b_i$ and $C^d_i$, while the transport rate $c^t$ is fixed. For this reaction network, a symmetry-reduced model of up to the second order can be described using $\hat{K}_{eq}=12$ differential equations regardless of the population size.

The goal of this case study is to study how cell-to-cell communication affects the variability in the abundance of $\mathrm{P}_i$. To quantify variability, we use two metrics: the first one is the coefficient of variation (CV) defined as 
\begin{equation} \label{cv}
    \mathrm{CV}[P_i] = \sqrt{\frac{\langle P_i^2 \rangle -\langle P_i\rangle ^2}{\langle P_i\rangle^2}},
\end{equation}
for any $i=1, \ldots, N$. The CV captures the expected variation in protein abundance inside single cells across \textit{different} populations. Due to the symmetry, we have $\mathrm{CV}[P_i] = \mathrm{CV}[P_k]$ for any $i$ and $k$. Furthermore, we define the pair variation (PV)
\begin{equation} \label{pv}
    \mathrm{PV}[P_i,P_k] = \sqrt{\frac{\langle (P_i-P_k)^2\rangle}{\langle P_i\rangle \langle P_k \rangle }},
\end{equation}
which captures the expected variation between two different cells $i$ and $k$ within the \textit{same} population. Note that in the absence of cell-cell communication, (\ref{cv}) and (\ref{pv}) are identical up to a scaling factor of $\sqrt{2}$. For a communicating population, however, this is not the case due to correlations between cells in the population.

\begin{figure}[h!]
\centering
\includegraphics[width=\columnwidth]{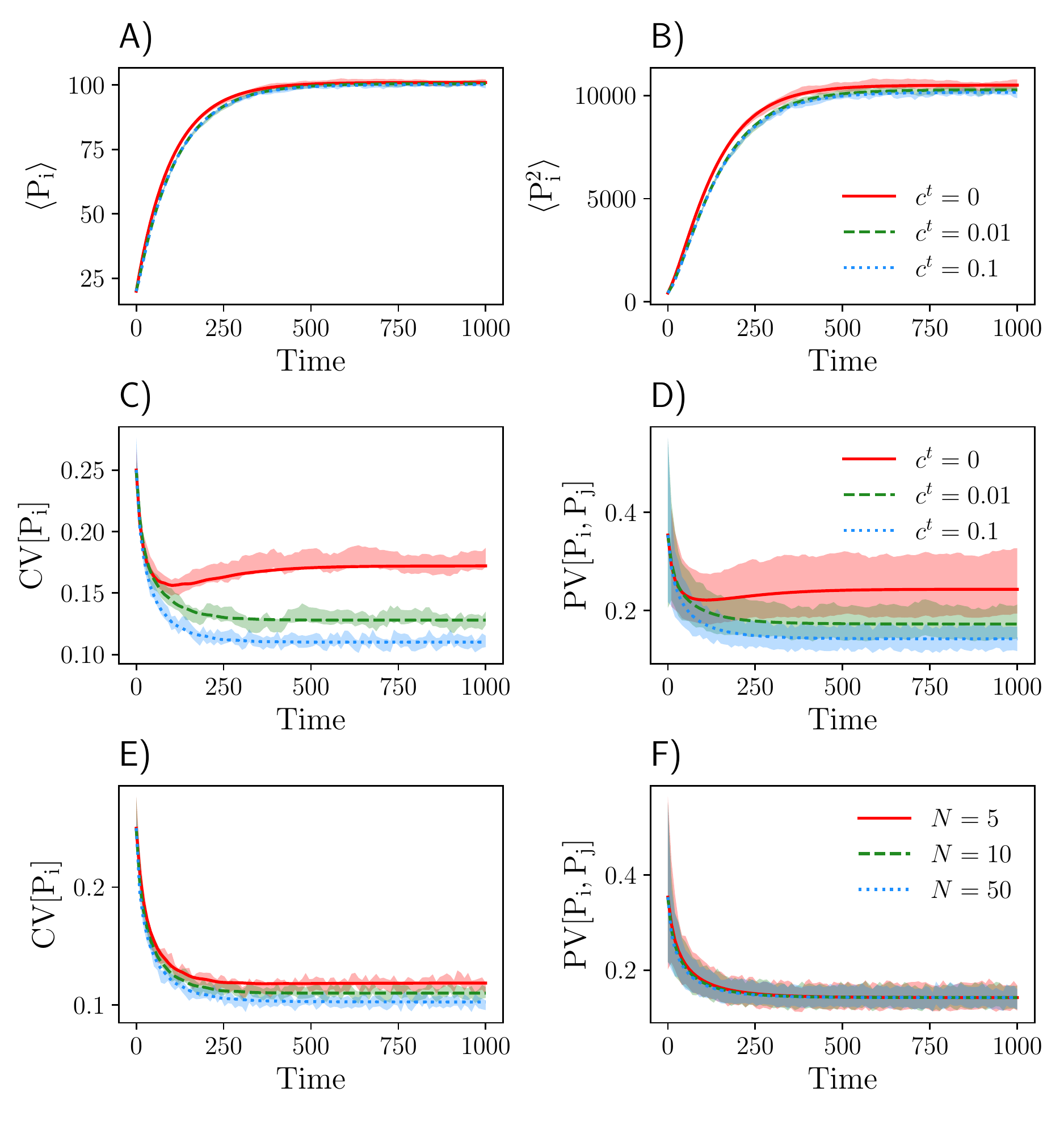}
\vspace*{-8mm}
\caption{Moment dynamics of the birth-death system generated by the symmetry-reduced model (Normal MA). First (A) and second order (B) moment dynamics of species $P_i$ with a population size of $N=10$ and $c^t=[0, 0.1, 0.1]$. CV[$P_i$] and PV[$P_i,P_j$] decrease as transport rate increases for a fixed population size $N=10$ (C and D). Increasing the population size $N=[5,10,50]$ for a fixed $c^t=0.1$ decreases the CV[$P_i$], but not PV[$P_i,P_j$] (E and F). Other parameters and initial conditions were set to $\langle C^b_i \rangle = 1$, $\mathrm{Var} [C^b_i] = 0.01$, $\langle C^d_i \rangle = 0.01$, $\mathrm{Var} [C^d_i] = 1e-6$, $\langle P_i(0) \rangle = 20$, $\mathrm{Var} [P_i(0)] = 25$, $\langle Q(0) \rangle = 0$, and $\mathrm{Var} [{Q}(0)]  = 0$. Shaded areas are bootstrapped 95\% confidence intervals (CI) from $1000$ SSA realizations.}
\label{fig:birthdeath}
\end{figure}

In Figs. \ref{fig:birthdeath}A and B, we compare the first and second order moments of species $P_i$ for the symmetry-reduced model of size $N=10$ with stochastic simulations and found a good agreement between the two approaches. In Figs. \ref{fig:birthdeath}C-F, we show the dependency of CV[$P_i$] and PV[$P_i$,$P_j$] as a function of the transport rate as well as the population size. The coefficient of variation CV[$P_i$] decreases with increasing transport rates and to some extent also with the population size. While the pair variation PV[$P_i$,$P_j$] shows a similar decrease with increasing transport rates, it seems to be independent of the population size.

\begin{figure}[h]
\centering
\includegraphics[width=\columnwidth]{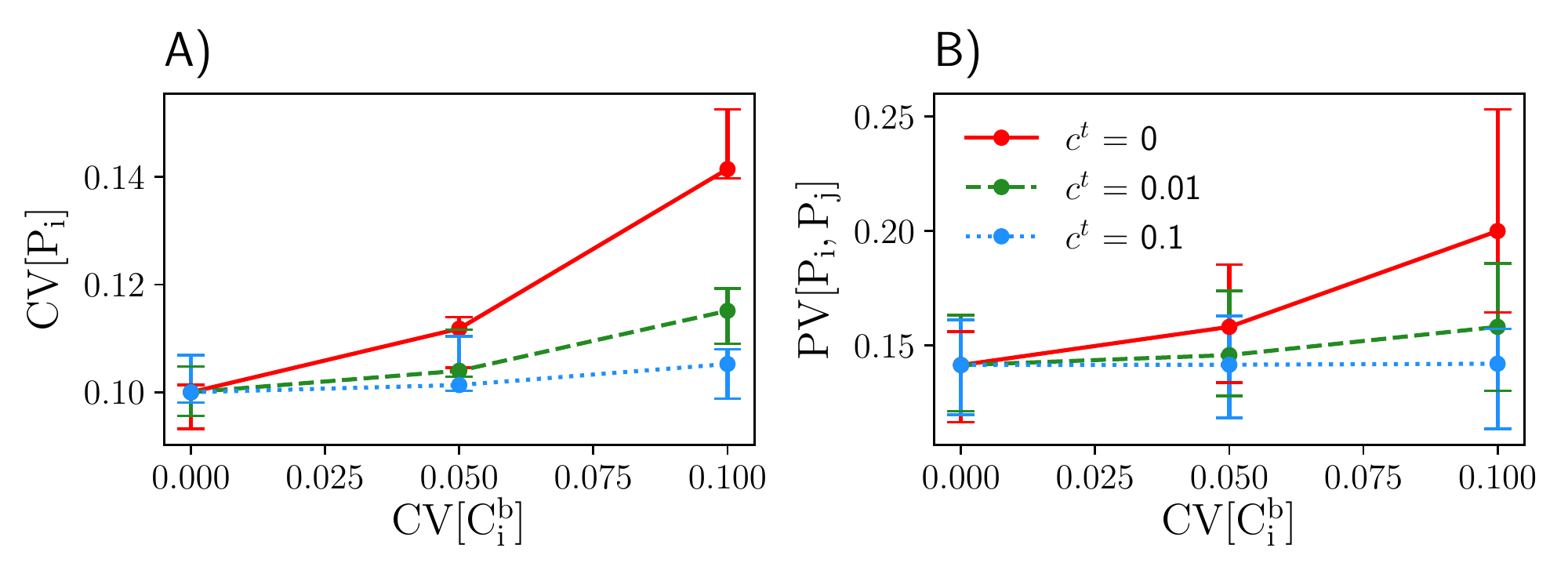}
\vspace*{-8mm}
\caption{Steady-state CV[$P_i$] (A) and PV[$P_i,P_j$] (B) for varying transport rates ($c^t$) and population heterogeneity (CV$[C^b_i]$). Other parameters and initial conditions were set to $\langle C^b_i \rangle = 1$, $c^d = 0.01$, $\langle P_i(0) \rangle = 20$, $\mathrm{Var} [P_i(0)] = 25$, $\langle Q(0) \rangle = 0$, $\mathrm{Var} [{Q}(0) ]= 0$, and $N=10$. Steady-state moments were determined from the reduced model with normal closure at $t=1000$. Error bars are bootstrapped 95\% CIs from $1000$ SSA realizations.}
\label{fig:birthdeath_ex}
\end{figure}

We also show how extrinsic variability changes the steady-state variability of $\mathrm{P}_i$ by increasing the CV of the birth rate parameter $C^b_i$, while keeping all other parameters constant. Both CV[$P_i$] and PV[$P_i$,$P_j$] increase with increasing extrinsic variability, whereas large transport rates can attenuate this effect due to spatial averaging of $P_i$ levels (Fig. \ref{fig:birthdeath_ex}).

\subsection{Autocatalytic circuit}
Next, we focus on an autocatalytic system defined by
\begin{equation} \label{autocatalytic}
\begin{gathered}
    \begin{aligned}[c]
    \emptyset \xrightarrow{C^b_i} \mathrm{A}_i \xrightarrow{C^d_i} \emptyset\\
    \mathrm{A}_i \xrightarrow{c^a} \mathrm{A}_i + \mathrm{A}_i
    \end{aligned}
    \qquad\qquad
    \begin{aligned}[c]
    \mathrm{A}_i \xleftrightharpoons[c^t]{c^t} \mathrm{B}.
    \end{aligned}
\end{gathered}
\end{equation}

\begin{figure}[h]
\centering
\includegraphics[width=\columnwidth]{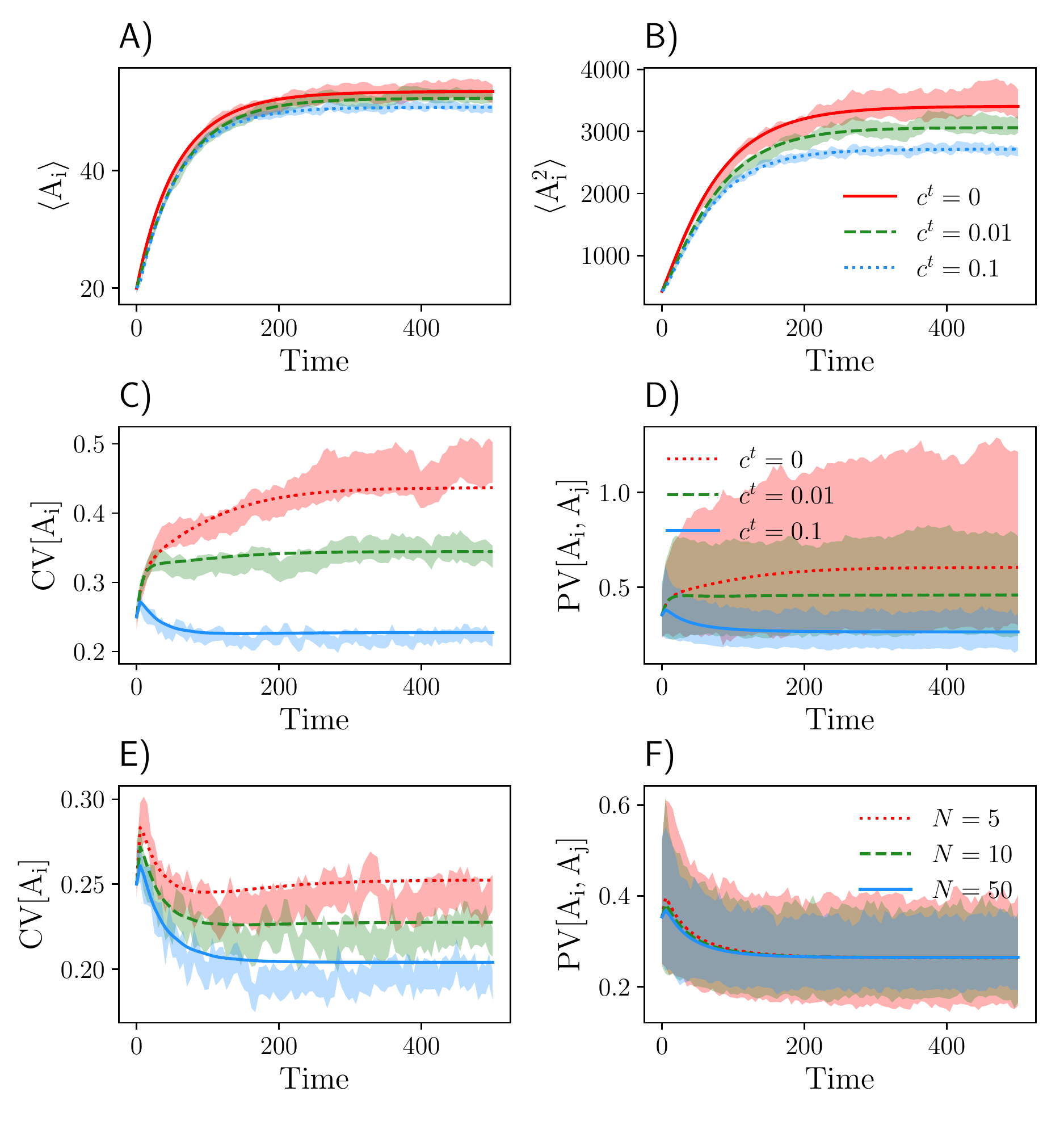}
\vspace*{-8mm}
\caption{Moment dynamics of the autocatalytic system generated by the symmetry-reduced model (Lognormal MA). First (A) and second order (B) moment dynamics of species $A_i$ for a population size of $N=10$ and $c^t=[0,0.01,0.1]$. CV[$A_i$] and PV[$A_i,A_j$] decrease as transport rate increases for a fixed population size $N=10$ (C and D). Increasing the population size from $N=[5,10,50]$ for a fixed $c^t=0.1$ decreases CV[$A_i$] (E), while PV[$A_i,A_j$] remains the same (F). Other parameters and initial conditions were set to $\langle C^b_i \rangle = 1$, $\mathrm{Var} [C^b_i] = 0.01$, $c^a = 0.08$, $\langle C^d_i \rangle = 0.1$, $\mathrm{Var}[C^d_i] = 1e-4$, $\langle A_i(0) \rangle = 20$, $\mathrm{Var}[ {A_i}(0)] = 25$, $\langle B(0) \rangle = 0$, and $\mathrm{Var}[ {B}(0) ] = 0$. Shaded areas are bootstrapped 95\% CIs from $1000$ SSA realizations.}
\label{fig:autocatalytic}
\end{figure}

Here we consider $C^b_i$ and $C^d_i$ to be randomly distributed across the population, while the autocatalytic rate $c^a$ and transport rate $c^t$ are fixed. In this example, the total number of reduced equations is $\hat{K}_{eq}=12$. To obtain a closed set of moments, we applied the lognormal closure and checked its accuracy by comparing it to Monte Carlo estimates of the moments calculated over $1000$ SSA realizations (Fig. \ref{fig:autocatalytic}). We generally found a good agreement between the MA and the SSA. Similar to the birth-death systems, we observe that both CV and PV of species $\mathrm{A}_i$ decrease with increasing transport rates. Increasing the population size does not affect the PV, but decreases its CV.

\begin{figure}[h!]
\centering
\includegraphics[width=\columnwidth]{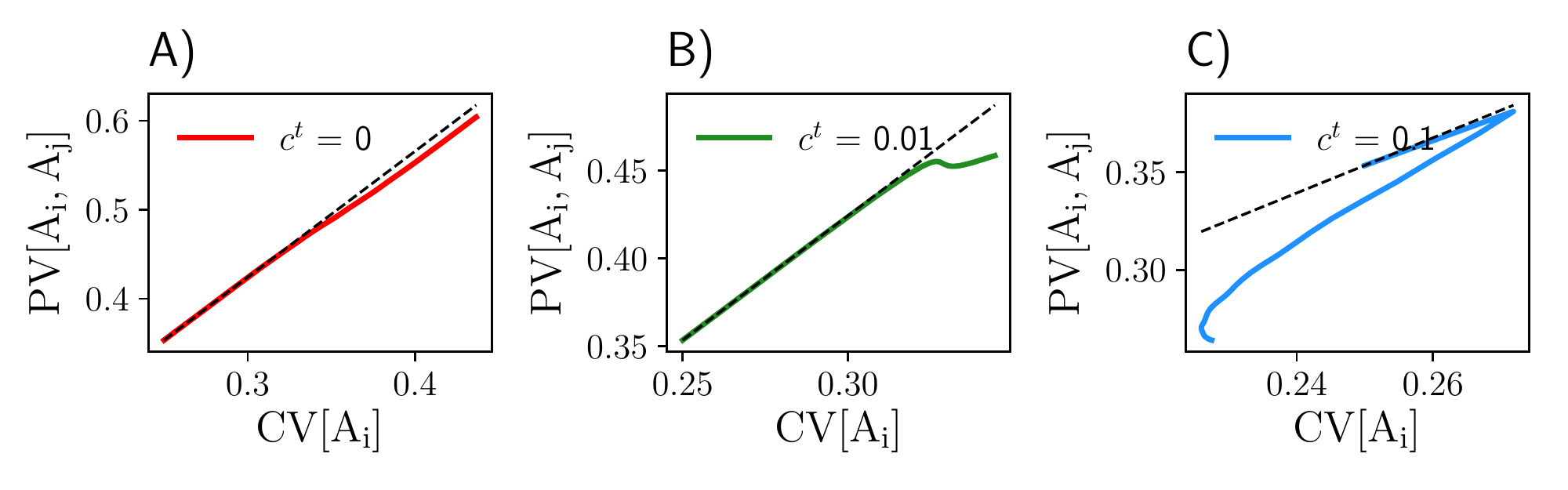}
\vspace*{-8mm}
\caption{CV[$A_i$] vs. PV[$A_i$,$A_j$] of the autocatalytic network at different transport rates $c^t=[0,0.01,0.1]$. CV[$A_i$] and PV[$A_i$,$A_j$] are related by a factor of $\sqrt{2}$ (diagonal) at $c^t=0$, while deviations from this scaling occur when $c^t > 0$. Values of CV[$A_i$] and PV[$A_i$,$A_j$] were taken from the time-courses corresponding to the case $N=10$ in Fig. \ref{fig:autocatalytic}.}
\label{fig:cv_vs_pv}
\end{figure}

To study the relationship between the CV and PV, we plotted CV[$A_i$] against PV[$A_i$,$A_j$] in Figure \ref{fig:cv_vs_pv}. As expected, CV and PV are related by a factor of $\sqrt{2}$ when $c^t = 0$. Slight deviations from this scaling are possible due to the approximations involved in the derivation of the moment equations. In the presence of communication, PV[$A_i$,$A_j$] drops below the $\sqrt{2}$ scaling law, indicating that the variability between cells in the same population is smaller than the variability of cells across different populations.

\subsection{Genetic feedback circuit} \label{feedback}
\begin{figure}[h!]
\centering
\includegraphics[width=\columnwidth]{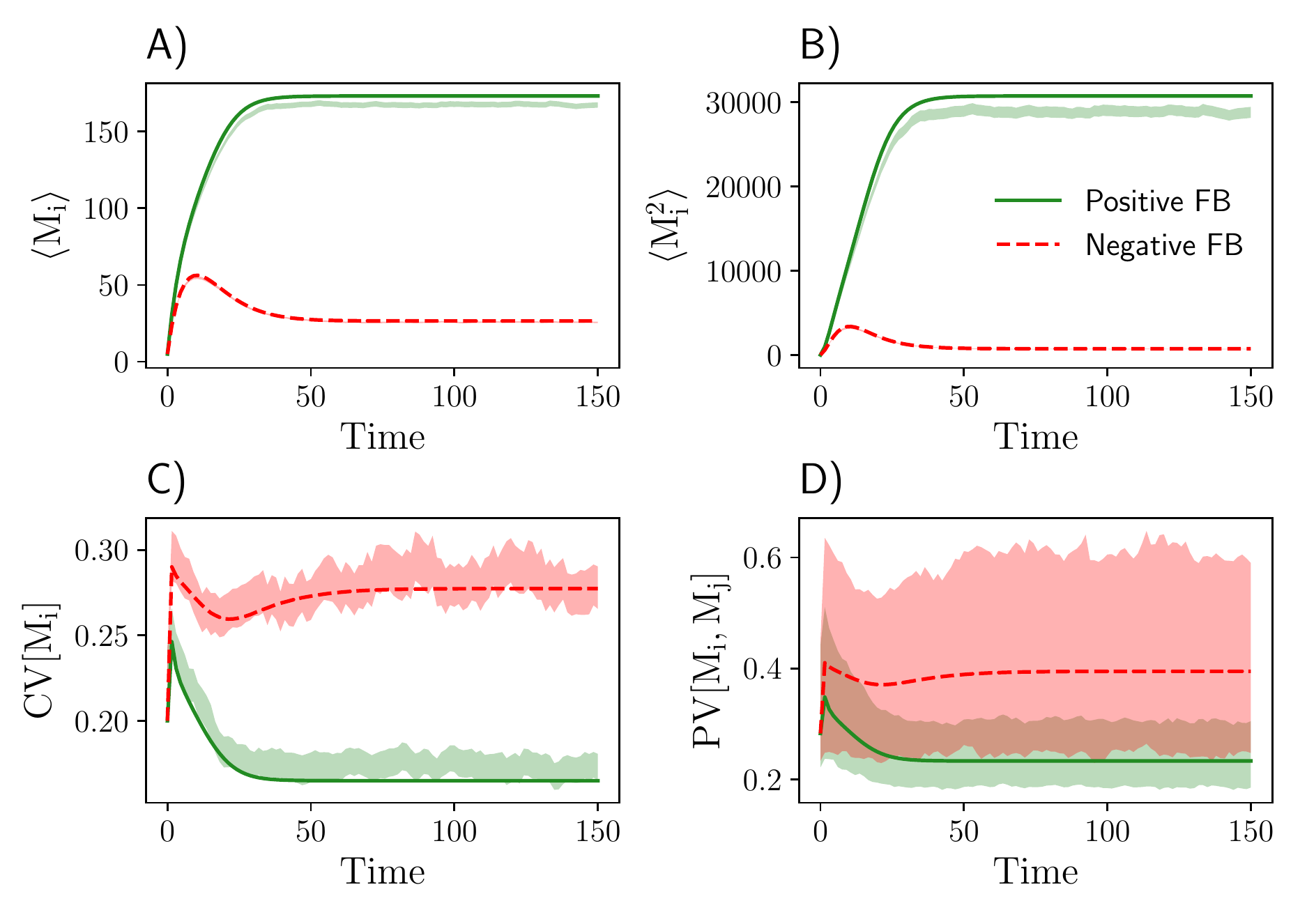}
\vspace*{-8mm}
\caption{Moment dynamics of the genetic feedback circuit generated by the symmetry-reduced model (Lognormal MA). First (A) and second order (B) moment dynamics of species $M_i$ for populations of $N=10$ cells with negative feedback ($c^f_i = 0.1$) and positive feedback ($c^f_i=1$) and their corresponding time evolution of CV[$M_i$] (C) and PV[$M_i,M_j$] (D). Reaction rate parameters and initial conditions were set to $\langle D_i(0) \rangle = 30$, $\mathrm{Var}[ {D_i}(0)] = 25$, $\langle M_i(0) \rangle = \langle P_i(0) \rangle = \langle DP_i(0) \rangle = 5$, $\mathrm{Var}[ {M_i}(0) ] = \mathrm{Var}[ {P_i}(0)] = \mathrm{Var}[{DP_i}(0)] = 1$, $\langle Q(0) \rangle = 5$, $\mathrm{Var}[{Q}(0)]= 0$, $\langle C^m_i\rangle = 0.5$, $\mathrm{Var}[{C^m_i}]= 0.01$, $\langle C^p_i\rangle = 0.05$, $\mathrm{Var}[C^p_i]= 2.5e-5$, $c^a=0.01$, $c^d=0.01$, $d^m=d_p=0.2$, and $c^t=0.8$. Shaded areas are bootstrapped 95\% CIs from $1000$ SSA realizations.}
\label{fig:feedback}
\end{figure}

Lastly, we tested the moment-based method using a larger system. In particular, we focus on a genetic feedback circuit given by
\begin{equation} \label{feedback}
\begin{gathered}
    \begin{aligned}[c]
    \mathrm{D}_i \xrightarrow{C^m_i} \mathrm{D}_i + \mathrm{M}_i \\
    \mathrm{M}_i \xrightarrow{C^p_i} \mathrm{M}_i + \mathrm{P}_i \\
    \mathrm{D}_i + \mathrm{P}_i \xleftrightharpoons[c^d]{c^a} \mathrm{DP}_i \\
    \mathrm{DP}_i \xrightarrow{c^f} \mathrm{DP}_i + \mathrm{M}_i
    \end{aligned}
    \qquad\quad
    \begin{aligned}[c]
    \mathrm{M}_i \xrightarrow{d^m} \emptyset \\
    \mathrm{P}_i \xrightarrow{d^p} \emptyset \\
    \mathrm{P}_i \xleftrightharpoons[c^t]{c^t} \mathrm{Q},
    \end{aligned}
\end{gathered}
\end{equation}
for $i=1, \ldots, N$. Here, we consider the reaction rate constants associated with transcription $C^m_i$ and translation $C^p_i$ to be randomly distributed across the population, while all other rate parameters are fixed. Transcriptional feedback of the gene circuit is mediated by the protein product ($\mathrm{P}$), which binds DNA ($\mathrm{D}$) to form a complex $\mathrm{DP}$. In the bound state, the gene can be transcribed with rate constant $c^f$. Depending on the ratio of the bound and unbound transcription rate, the feedback mechanism can either enhance ($c^f > \langle C^m_i \rangle$) or inhibit ($c^f < \langle C^m_i \rangle$) gene expression. We consider a population of $N=10$ cells and applied a lognormal closure to solve for the reduced moment dynamics. The total number of reduced equations for this system is $\hat{K}_{eq}=48$. Fig. \ref{fig:feedback} shows that the time evolution of the mRNA moments obtained from the symmetry-reduced model and SSA are very similar to each other.  

To analyze the computational efficiency of the moment-based approach, we recorded run times for the positive feedback gene circuit for different population sizes $N$ and compared it against the SSA (Table \ref{compute_time}). As expected, the run time for the moment-based model does not increase with $N$ while the run time associated with the SSA increases by orders of magnitude.

\begin{table}[h!]
    \centering
    \footnotesize
    \caption{Run times (s) of the reduced moment-based approach and SSA for the positive feedback gene circuit*.}
    \begin{tabular}{ccc}
    \hline
    \noalign{\vskip 1mm} 
    $N$ & Reduced moments & SSA (1 sample path) \\
    \noalign{\vskip 1mm}
    \hline
    \noalign{\vskip 1mm}  
    10 & $0.068\pm0.005$ & $0.142\pm0.004$\\ 
    \noalign{\vskip 1mm}
    100 & $0.065\pm0.004$ & $12.300\pm0.123$\\ 
    \noalign{\vskip 1mm}
    1000 & $0.064\pm0.005$ & $2569.20\pm52.93$\\
    \noalign{\vskip 1mm}
    \hline
    \noalign{\vskip 1mm}
    \end{tabular}
    \\ *Mean and standard deviation from 10 replicates.
    \label{compute_time}
\end{table}

\section{Conclusions}
We presented and analyzed an efficient moment-based approach to study noise in heterogeneous populations of communicating cells. 
By exploiting certain symmetries of the underlying moment dynamics \cite{batmanov2012}, the number of differential equations needed to describe the population could be strongly reduced. Indeed, the dimensionality of the reduced model is independent of the number of considered cells, which enables the analysis of regulatory processes in large populations. In principle, the method applies to arbitrary intracellular reaction networks, although its accuracy relies on the availability of suitable moment closure functions. In the present work, we tested the approach using three different case studies, where it accurately captured first and second order moments using normal and lognormal closure functions. Our analyses show that moment-based approximations provide an effective way to study how biochemical noise and population heterogeneity affect the collective dynamics of communicating cells.

\bibliographystyle{unsrt}
\bibliography{bibliography}

                                                         
\end{document}